# 10-Hertz quantum light source generation on the cesium D$_2$ line using single photon modulation

Guan-Hua Zuo[1], Yu-Chi Zhang[2], Gang Li[1] Peng-Fei Zhang[1], Peng-Fei Yang[1], Yan-Qiang Guo[4,†], Shi-Yao Zhu[3]
Tian-Cai Zhang[1,‡]

[1]*State Key Laboratory of Quantum Optics and Quantum Optics Devices, Collaborative Innovation Center of Extreme Optics, Institute of Opto-Electronics, Shanxi University, Taiyuan 030006, China*
[2]*College of Physics and Electronic Engineering, Shanxi University, Taiyuan 030006, China*
[3]*Department of Physics, Zhejiang University, Hangzhou 310027, China*
[4]*Key Laboratory of Advanced Transducers and Intelligent Control System, Ministry of Education and Shanxi Province, College of Physics and Optoelectronics, Taiyuan University of Technology, Taiyuan 030024, China*
*Corresponding author. E-mail:* †*guoyanqiang@tyut.edu.cn,* ‡*tczhang@sxu.edu.cn*

Generation of quantum light source is a promising technique to overcome the standard quantum limit in precision measurement. Here, we demonstrate an experimental generation of quadrature squeezing resonating on the cesium D$_2$ line down to 10 Hz for the first time. The maximum squeezing in audio frequency band is 5.57 dB. Moreover, we have presented a single-photon modulation locking to control the squeezing angle, while effectively suppressing the influence of laser noise on low-frequency squeezing. The whole system operates steadily for hours. The generated low-frequency quantum light source can be applied in quantum metrology, light-matter interaction investigation and quantum memory in the audio frequency band and even below.

## 1 Introduction

Quantum light source plays an important role in many areas of continuous-variable quantum physics since it first generation in 1985 [1]. With its excellent quantum features, squeezed light, especially which resonates on atomic transition, plays an important role in quantum communication [2–5], quantum storage [6,7], light-atom interaction investigation [8–10], and precision measurement [11–14]. An optical parametric amplifier (OPA) or optical parametric oscillator working subthreshold is a typical device for the generation of squeezed light [15]. Many experiments have produced squeezing at megahertz or hundreds of kilohertz frequencies since laser noise and various technical noises are very close to the shot-noise limit in these frequency bands [16,17]. However, in some practical applications, such as gravitational wave (GW) detection [18], biological magnetic measurement [19,20], and the interaction survey between light and atomic media [21], the analysis frequency is tens of hertz or even lower. With the development of precise measurement, the vacuum fluctuation of light has become the final measurement limitation [22,23]. Therefore, the preparation of squeezed light at low frequencies is needed to improve the measurement sensitivity beyond the classical limitation [24]. In fact, a GW observatory operating beyond the standard quantum limit (SQL) has been realized experimentally [25,26], and the sensitivity of the Laser Interferometer GW Observatory (LIGO) interferometers above 50 Hz is improved by up to 3 dB using squeezed states [27]. Squeezed-light-enhanced atomic magnetometers have been demonstrated recently [28,29]. In 2010, Wolfgramm et al. have employed squeezed light as a probe for an atomic magnetometer and demonstrated a sensitivity improvement of magnetic measurement [30]. However, the above quantum-enhanced atomic magnetometer was fulfilled over tens of kilohertz band where squeezing is relatively easy to be prepared compared to the audio frequency band, especially at tens of hertz or even lower. Therefore, it is particularly important to prepare squeezed light in low-frequency band.

Observing squeezing in a low-frequency band is difficult because there is considerable noise. There are two kinds of noise: one is various technical noises, such as mechanical vibration, parasitic interference, and beam pointing noise [31,32]; and the other is the laser noise of the control light, such as the pump, probe, and lock beams, of the OPA coupling in the squeezed light [33]. These noises can easily corrupt the squeezing level in a low frequency band or even completely overwhelm the squeezing. In addition, a balanced homodyne detector (HD) with low electronic noise and a high common-mode rejection ratio (CMRR) in a low frequency band is also required [34].

Many researchers have conducted extensive and deep studies on the squeezed light generation in a low frequency band [35–38]. A quantum noise locking method has been developed and a squeezed vacuum state down to 280 Hz at 1064 nm was obtained [33,39]. However, the pump phase was not controlled, so the squeezing angle was free to evolve. Valbruch et al. successfully generated a squeezed vacuum state at 1064 nm down to 1 Hz by eliminating parasitic interference and laser noise [34]. Whereas the low-frequency squeezing shown above were all at a GW detection wavelength of 1064 nm, and to interact with the atomic media, the wavelengths resonant on the atomic transition lines need to be chosen. For the cesium (Cs)

D$_2$ line, because the wavelength of 852 nm is considerably shorter than 1064 nm, problems with absorption and heating effects arise. The absorption of periodically poled KTiOPO$_4$ (PPKTP) crystal at 852 nm is much larger than that at 1064 nm, which is approximately 1%/cm at 852 nm and approximately 10%/cm at 426 nm, while it is approximately 0.02%/cm at 1064 nm [40]. The absorption loss results in the degradation of the squeezing level, and the heating effect leads to the instability of the cavity length stabilizing and pump phase controlling, which limits the squeezing band towards low analysis frequencies. Accordingly, the analysis frequencies of the generated squeezed light on the atomic transition lines reported at present are all above the kilohertz band [41–45].

Controlling the squeezing angle is the basic problem for squeezed light applications, and it is necessary to precisely control the squeezing angle in many applications. To improve the sensitivity of GW detection in the Advanced LIGO detector, a squeezed state with tuning of the squeezing angle is required [46,47]. Moreover, in the quantum-enhanced Mach-Zehnder interferometer, to improve the signal-to-noise ratio of the measurement, a squeezed state with squeezing angle of $\pi/2$ is employed [48]. It is also found that the anti-bunching effect of the squeezed coherent state can be produced by controlling the squeezing angle [49]. Furthermore, polarization-squeezed light can be achieved by combining a dim quadrature squeezed beam with a squeezing angle of 0 or $\pi/2$ with a bright coherent beam or another quadrature squeezed beam on a polarizing beam splitter (PBS) [50]. Common squeezing angle control schemes rely on the injection of a weak and phase-modulated probe light at the fundamental frequency into the squeezed light source. It has been shown that even the lowest probe power will introduce large amounts of classical noise, such as probe light noise, cavity detuning, and pump light noise at audio frequencies and below, and squeezing can no longer be achieved [33]. In 2006, Schnabel et al. proposed a coherent control method to control the squeezing angle of the squeezed vacuum state and the phase relative to the local light [51]. This method requires frequency-shifted light to stabilize the cavity length and another frequency-shifted light to control the pump phase and the local phase. Therefore, two separate lasers had to be used, and the control system was very complex.

In this letter, we demonstrate a generation of broadband quadrature squeezed light by a subthreshold OPA down to 10 Hz resonant on the Cs D$_2$ line with a maximum squeezing of 5.57 dB, which is the lowest squeezing band on the atomic transition line reported to date. Moreover, we have presented a single-photon modulation locking (SML) to control the squeezing angle, while effectively suppressing the influence of laser noise on low-frequency squeezing. The generated low-frequency quantum light source on the atomic transition line can be applied in quantum metrology and quantum information storage in electromagnetically induced transparency (EIT) media.

## 2 Experiment setup

Many obstacles limit squeezing to audio frequencies and below. At these frequency bands, there are many noise sources. The most major source is the laser noise, as a Ti:sapphire laser has high background noise at low frequencies, which will be easily coupled into the squeezed light. To avoid noise coupling, we use a frequency-shifted and phase-modulated field to stabilize the cavity length and reduce the probe power injected into the OPA as much as possible.

The experimental system is shown in Fig. 1. The main laser is a low-noise tuneable Ti:sapphire laser (Msquare Co.) in which the frequency is locked to the Cs D$_2$ line at 852.3 nm by the polarization spectroscopy method. Squeezed light is generated from a subthreshold OPA, which is pumped by a 426.2 nm pump light generated by a second harmonic generator (SHG). The cavity of the OPA has a bow-tie-type ring configuration consisting of two spherical mirrors (radius of curvature of 50 mm) and two flat mirrors. One of the flat mirrors is the output coupler, and the transmittance is T = 0.11. The cavity round trip length is 407 mm. A 10-mm-long type-I PPKTP crystal (Raicol Crystals) is placed between the two spherical mirrors. The crystal is specially anti-reflecting coated to reduce the intracavity losses by placing it in a copper-made oven and stabilizing the temperature to 46.5 °C for optimization. The beam waist size inside the crystal is 22.7 μm. To stabilize the cavity length without bringing the noise of the lock light into the generated squeezed light, a semiconductor laser (Waviclelaser Co.) locked to the Cs D$_1$ line at 894.6 nm by the saturation absorption spectroscopy (SAS) method is employed. We used an avalanche photodiode (APD, C30659-900-R5B) to record the faint lock light signal. By tuning the modulation frequency, the sideband frequency of the 894.6 nm lock light and the 852.3 nm probe light are both simultaneously made to resonate to the OPA cavity and locked using the Pound-Drever-Hall technique. Various methods are used for mitigating parasitic interference in HD detection. Optical components with minimal surface roughness for reducing the amount of scattering are used. Beam dumps are carefully placed to dump the scattered photons, and the phase fluctuations in the scattered fields are decreased by reducing the vibration in the optical system. And a triangular ring cavity mode cleaner (MC) is built in the local path to decrease the beam pointing noise while increasing the interference visibility of the homodyne. Moreover, the stable cavity design and usage of ultrastable mirror mounts help to increase mechanical stability. Finally, the whole optical system is built on an air-floating and highly stable optical platform, and shielded by a sound-proof cover to mitigate the sound variation of the surroundings. To detect squeezing in the low frequency band, a homemade HD is utilized. Its average CMRR from 10 Hz to 300 kHz is approximately 55 dB, and the electric dark noise is 7 dB below the shot noise of 2 mW local light at 10 Hz and is more than 10 dB above 100 Hz.

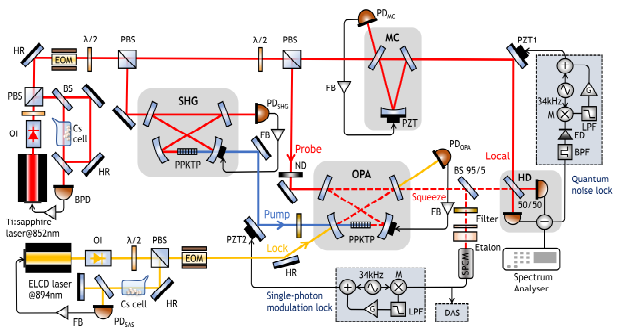

Fig. 1 Schematic of the experimental setup. OI: optical isolator, HR: high-reflectivity mirror, λ/2: half-wave plate, BS: beam splitter, ND: neutral-density filter, EOM: electro-optical modulator, FB: feedback control circuit, PD: photodetector, BPD: balanced photodetector, PZT: piezoelectric transducer, SPCM: single-photon counting module, DAS: data acquisition system. The dashed black lines indicate the electronics for the phase control process. G: amplifier circuit (gain), M: mixing circuit, BPF: bandpass filter, ED: envelope detector, LPF: low-pass filter.

To control the local phase in HD detection, a quantum noise locking method is used. The error signal is generated by dithering the local phase with 34 kHz and then demodulating the noise power of the HD. The noise power of the HD is divided into two parts. One part is detected using a spectrum analyser (Hewlett Packard-8590D, zero span at 2 MHz, resolution bandwidth (RBW) = 300 kHz, video bandwidth (VBW) = 30 kHz) and demodulated with a lock-in amplifier (Stanford Research System-SR830) with a modulation frequency of 34 kHz and amplitude of 0.23 V and then fed back to PZT1. By switching the phase in the proportional-integral-derivative controller, either a squeezed phase or an anti-squeezed phase locking can be achieved. Another part is measured by a low-frequency spectrum analyser (Rohde& Schwarz, FSV-4). When no probe light is seeded, we measure squeezing traces at 5 kHz as the homodyne phase is varied. The results are shown in Fig. 2. A vacuum squeezing of 5.70 dB is obtained, and the anti-squeezing is 13.68 dB. The observed squeezing $R_-$ and anti-squeezing $R_+$ are obtained as follows [52]:

$$R_\pm = 1 \pm \eta \xi^2 \zeta \rho \frac{4x}{(1 \mp x)^2 + 4\Omega^2}, \quad (1)$$

where $R_\pm$ is the generated squeezing/anti-squeezing without taking into account the phase fluctuation, with quantum efficiency of the photodiode $\eta = 0.990$, the homodyne visibility $\xi = 0.985$, propagation efficiency $\zeta = 0.912$, for which the lower propagation efficiency is mainly due to the splitting ratio of 95/5 BS, and the escape efficiency of the cavity $\rho = T/(T+L) = 0.879$, where $T$ is transmittance of the output coupler, and $L$ is the intracavity losses. $x = (P/P_{th})^{1/2}$ is the normalized pump power, with $P$ the pump power, and $P_{th} = 165 mW$ the oscillation threshold of the OPA. $\Omega = f/\gamma$ is the normalized frequency, with $f$ the measured frequency of the spectrum analyser, $\gamma = c(T+L)/l$ the OPA cavity decay rate, and $l$ the round trip length of the cavity. The expected squeezing at 5 kHz is 6.01 dB.

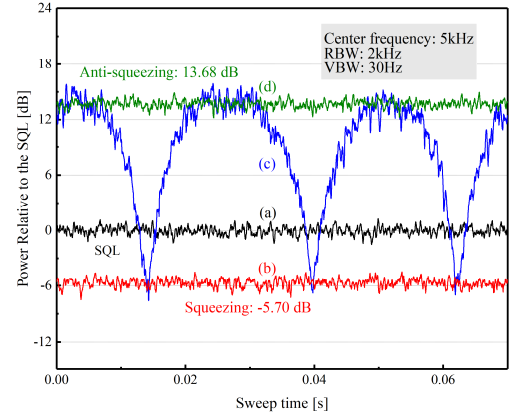

Fig. 2 Measured noise power as the phase of the homodyne is scanned with pump power of 100 mW and local power of 2 mW. (a) SQL; (b) Local phase is locked at the squeezed phase; (c) The local phase is scanned; (d) Local phase is locked at the anti-squeezed phase. These levels are normalized to make the SQL 0 dB. The measurement frequency is 5 kHz with RBW = 2 kHz, and VBW = 30 Hz. Electronic noise is subtracted from the data. Traces (a), (b), and (d) are averaged from 10 measurements. The observed squeezing/anti-squeezing are -5.70/+13.68 dB, respectively.

## 3 Generation of quadrature squeezing down to 10 Hz

To observe squeezing in low frequency band, the probe power should be decreased as much as possible [33]. Therefore, we reduce the probe power injected into the OPA to the single-photon level and propose a single-photon modulation method to control the squeezing angle, which can not only achieve control of the squeezing angle but also avoid coupling the laser noise into the squeezed light.

We use a 95/5 BS to divide a small portion of the generated squeezed light to control the squeezing angle. The probe light is injected into the OPA from a highly reflective mirror with a reflectivity greater than 99.99%. When the probe power is 1 nW, the average photon number of the probe light injected into the cavity is only 0.076. Such weak light requires the use of a SPCM to detect the signal. Moreover, in the type-I parametric down-conversion, except for the fundamental frequency $\omega_0$, some other pairs of down-conversion modes with nondegenerate frequencies of $\omega_{\pm m} = \omega_0 \pm m\Omega_{opo}$ may also simultaneously resonate to the cavity, with $\Omega_{opo}$ as the free spectral range of the cavity and $m$=1, 2, …, $n$. To detect the interference signal between the photons generated by parametric down-conversion and the injected probe light distinctly, it is necessary to filter out the frequency modes different from the fundamental frequency. Therefore, two etalons are employed to filter out the unwanted modes before the squeezed light enters the SPCM. The lengths of the etalons are 3 mm and 11 mm. An 852 nm bandpass interference filter is used to remove other wavelength lights. By scanning the pump phase, the interference signal generated by the phase-sensitive parametric process is obtained. The output of the SPCM is divided into two parts: one enters a data acquisition system (QuTag) to acquire the photon count rate, and the other is demodulated with a lock-in amplifier (SR830) with a modulation frequency of 33.5 kHz and amplitude of 0.45 V and then fed back to PZT2. The control electronics for phase control are indicated by the black dashed lines in Fig. 1.

When the probe power is 3 nW (the average photon number of probe light injected into the OPA is 0.23) and the pump power is 100 mW, by locking the pump phase to 0 (parametric amplification with a squeezing angle of π/2), the count rate is shown in Fig.3 (i). The oscillation signal is the interference signal of the photons when the pump phase is scanned, and the back part signal is the result of phase locking. The count rate is 1.95 MHz with only the pump light, which comes from the photons generated by the parametric down-conversion and the background counts (less than 1 kHz). When there is only probe light in the cavity, the count rate is 90 kHz. The bin width of the data acquisition is 20 ms. Similarly, when the probe power is 4.5 nW and the pump power is 90 mW, by locking the pump phase to π (parametric deamplification with a squeezing angle of 0), the count rate is shown in Fig. 3 (ii). The phase fluctuation within 1000 s of phase locking to 0 and π is 26.1 mrad and 11.7 mrad, respectively. The experimental results show that the squeezing angle control at the single-photon level can be well realized by the SML.

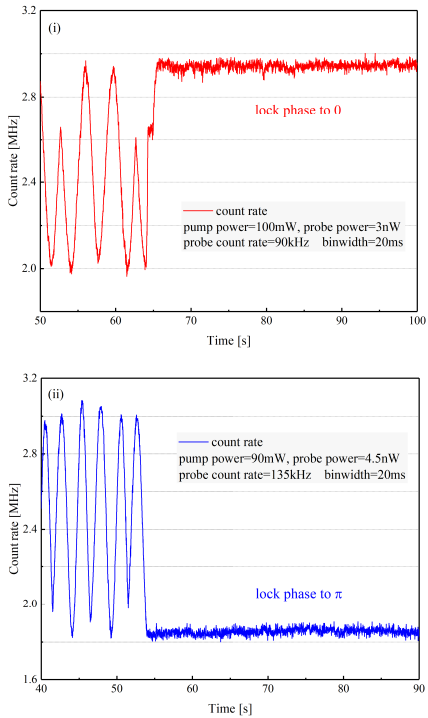

**Fig. 3** Experimental results of the pump phase controlled by the SML. (i) shows the phase locking to 0, corresponding to the parametric amplification with a squeezing angle of π/2, and (ii) shows the phase locking to π, corresponding to the parametric deamplification with a squeezing angle of 0. The bin width of the data acquisition is 20 ms.

With the help of pump phase control, squeezing angle-controlled quadrature squeezing is generated in the experiment. As the squeezing angle is locked to 0 and the local phase is locked by the quantum noise locking in the HD, quadrature squeezed light is observed experimentally. The squeezing spectrum between 10 Hz and 300 kHz is shown in Fig. 4. The probe power is 3 nW, the local power is 2 mW, and the pump power is 100 mW. The trace (a) is the SQL. The quadrature squeezing is the trace (b). The trace (c) is the anti-squeezing of 13.80 dB above the SQL. The electric dark noise is subtracted from the data. As a result, we obtain a broadband quadrature squeezed light down to 10 Hz. In addition, trace (b) shows that although the probe power is extremely low and the average photon number of probe light injected into the cavity is 0.23, the squeezing at approximately 10 Hz is degraded due to the roll-up laser noise and the imperfect CMRR of the HD in this frequency band. The peaks at 160 Hz and 25 Hz are the extra noise in the electric circuits, and the peaks at 34 kHz and 238 kHz are the modulation frequencies of the lock-in amplifier and its harmonic. The peaks at 30 Hz and 50 Hz are due to the electric noise of the SML electric circuits. For a significant portion of the shown spectrum, the maximum quadrature squeezing of 5.57 dB is obtained. In addition, the squeezing is slightly smaller than the vacuum squeezing of 5.70 dB due to the phase fluctuation introduced by the SML. Considering the phase fluctuation with a root-mean-square (RMS) of $\tilde{\theta}$, the observed squeezing $R'_-$ and anti-squeezing $R'_+$ are obtained as follows [52]:

$$R'_\pm = R_\pm \cos^2 \tilde{\theta} + R_\mp \sin^2 \tilde{\theta}. \quad (2)$$

where $R_\pm$ is the generated squeezing/anti-squeezing of the squeezed vacuum light. According to the results of the squeezing and anti-squeezing, the calculated phase fluctuation of the SML is 18.0 mrad, which agrees with the result obtained by the above experimental count rate.

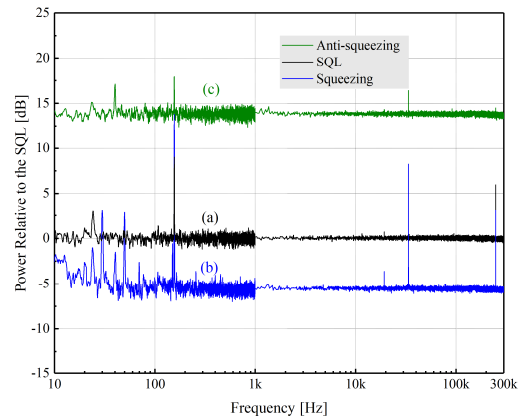

**Fig. 4** Measured noise spectrum for the (a) SQL, (b) quadrature squeezed light, and (c) anti-squeezed light. All traces are pieced together from two FFT frequency windows: 10 Hz-1 kHz and 1 kHz-300 kHz with RBW = VBW = 1 and 2 Hz, respectively. These levels are normalized to make the SQL 0 dB. Each point is the averaged RMS value of 200 and 100 measurements in the respective ranges. The electronic dark noise has been subtracted from the data.

To demonstrate squeezing in the 10 Hertz band, we used the zero span mode of the spectrum analyser to obtain a noise spectrum at 10 Hz and 70 Hz with a scan time of 10 s. The squeezing angle is locked to 0, the pump power is 100 mW, and the local power is 2 mW. As shown in Fig. 5, where Fig. 5 (i) is the result at 10 Hz and (ii) is at 70 Hz. The trace (a) is the SQL, the trace (b) is the observed quadrature squeezing, and the trace (c) is the observed anti-squeezing. Each measurement point is the averaged RMS value of 400 measurements. The observed squeezing is 2.85±0.41 dB at 10 Hz and 5.64±0.39 dB at 70 Hz. We note that each measurement trace in Fig. 5 lasts for more than an hour, thereby demonstrating the long-term stability of our system. To further evaluate the long-term stability of the system, the noise of the quadrature squeezed light at 10 kHz is recorded continuously for 1 hour. The fluctuation of the observed squeezing is ±0.17 dB over 1 hour. The experimental results show that the whole system can operate stably for a long time.

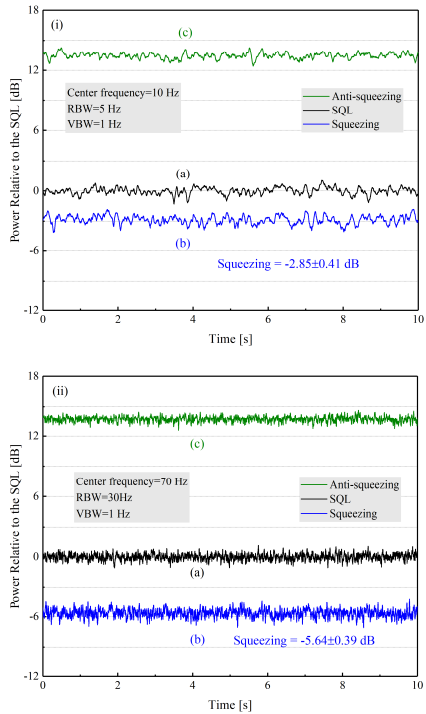

**Fig. 5** Observed noise power levels in zero span mode. (a) SQL. (b) Quadrature squeezing. (c) Anti-squeezing. These levels are normalized to make the SQL 0 dB. (i) Observed noise power at 10 Hz with RBW = 5 Hz and VBW =1 Hz. The observed squeezing is 2.85±0.41 dB. (ii) The results at 70 Hz with RBW = 30 Hz and VBW =1 Hz. The observed squeezing is 5.64±0.39 dB. All results are RMS averaged from 400 measurements and electronic noise is subtracted from the data.

## 4  Conclusion

In this paper, we have reported the generation of quantum light source of quadrature squeezed light down to 10 Hz resonant on the Cs $D_2$ line. The maximum squeezing of 5.57 dB in audio frequency band is obtained. In addition, we have presented a single-photon modulation locking to control the squeezing angle while effectively avoiding the influence of laser noise on low-frequency squeezing. The above experimental results benefit from the SML and the effective reduction of various technical noise in audio frequency band and below. The whole system can operate stably for hours, and the fluctuation of the observed squeezing at 10 kHz over 1 hour is ±0.17 dB. Two other methods are used to maintain the OPA operation: a frequency-shifted locking beam and the quantum noise locking method.

The generated quantum light source on the atomic transition line will be applied in quantum metrology, light-matter interaction research and quantum information storage. Squeezed states in a low frequency band can be used to create entanglement between light and atoms to push a light-atom interferometer beyond the SQL [53]. And the generated quantum light source and the presented SML scheme for squeezing angle control can be used in many practical applications that demand low frequency measurements, such as Cs atomic magnetometer [30], quantum memories based on EIT [54], and GW detection [55].

**Acknowledgements** This work was supported by National Natural Science Foundation of China (NSFC) (U21A6006, 11634008, 61875147 and 62175176) and National Key Research and Development Program of China (2017YFA0304502).